# Crowdsourced data indicates broadband has a positive impact on local business creation

Yifeng Philip Chen[a], Edward J. Oughton[b,*], Jakub Zagdanski[c], Maggie Mo Jia[a,d], Pete Tyler[a]

[a]*University of Cambridge, Cambridge, UK*

[b]*George Mason University, Fairfax, VA, United States*

[c]*GSMA Intelligence, London, UK*

[d]*Bader College, Queen's University, East Sussex, UK*

\* Corresponding author. E-mail address: eoughton@gmu.edu (E.J. Oughton).

ABSTRACT

Broadband connectivity is regarded as generally having a positive macroeconomic effect, but we lack evidence as to how it affects key economic activity metrics, such as firm creation, at a very local level. This analysis models the impact of broadband Next Generation Access (NGA) on new business creation at the local level over the 2011-2015 period in England, United Kingdom, using high-resolution panel data. After controlling for a range of factors, we find that faster broadband speeds brought by NGA technologies have a positive effect on the rate of business growth. We find that in England between 2011-2015, on average a one percentage increase in download speeds is associated with a 0.0574 percentage point increase in the annual growth rate of business establishments. The primary hypothesised mechanism behind the estimated relationship is the enabling effect that faster broadband speeds have on innovative business models based on new digital technologies and services. Entrepreneurs either sought appropriate locations that offer high quality broadband infrastructure (contributing to new business establishment growth), or potentially enjoyed a competitive advantage (resulting in a higher survival rate). The findings of this study suggest that aspiring to reach universal high-capacity broadband connectivity is economically desirable, especially as the costs of delivering such service decline.

**Declaration of Competing Interest**

The authors declare that they have no known competing financial interests or personal relationships that could have appeared to influence the work reported in this paper.

**Acknowledgements**

This work was supported by the UK Engineering and Physical Science Research Council programme grant entitled Multi-scale Infrastructure systems Analytics (EP/N017064/1). The authors would like to gratefully acknowledge the support of Janusz Jezowicz from Broadband Speed Checker for the provision of data. We also would like to thank the anonymous reviewers of the paper, and the support of editorial staff. There are no competing interests.

1. INTRODUCTION

Infrastructure is high in priority among policy makers to help drive economic growth and productivity (Glass and Tardiff, 2019; David, 2019; Hall et al., 2016a; Hall et al., 2016b), with many studies (e.g., Nadiri et al., 2018) suggesting that they contribute positively to the economy at the macroeconomic level. While basic broadband infrastructure access is known to be a necessary ingredient to enable local economic development (Ford and Koutsky, 2005; Kolko, 2010), there is debate around the level of influence high-speed broadband has on various economic metrics such as economic growth, employment, and business creation. The effects of broadband infrastructure at the very local level remain relatively overlooked compared to studies at more aggregated geographical levels (Holt and Jamison, 2009). This is particularly the case for the deployment of new broadband technologies and whether their availability enable or constrain business creation at the very local level. New types of digital infrastructure are a key reason for the rapid growth in digital content, applications and services driven by the digital entrepreneurial ecosystem (Sussan and Acs, 2017), supporting the idea that improved broadband leads to new firm creation.

The analysis presented in this article seeks to understand more about how the provision of broadband affects net business creation at the very local level using high-resolution spatial broadband data in England. The article defines business creation as the formation of new local business units, which have the potential to provide both additional economic value and employment to a local economy. The study focuses on the deployment of Next Generation Access (NGA), which is a highly important topic because of the significant emphasis given to these broadband technologies in telecommunications policy over the past decade (Feijóo et al., 2011). NGA is indicative of premises having access to at least Fibre-To-The-Cabinet (FTTC) which uses copper wire for only the final premises connection, or the cable technology Data-Over-Cable-Service-Interface-Specification (DOCSIS) 3.0 which uses coaxial cable for only the final premises connection. It also encompasses potentially higher capacity technologies such as Fibre-To-The-Premises where the connection is entirely fibre optic cable. Thus, all these technologies exceed the maximum 24 Mbps provided by ADSL2+. Using annual data from 2011 to 2015, a panel data analysis is undertaken of net business establishment growth in England for 6,791 Middle Super Output Areas (MSOAs).[1]

Growth in new businesses following broadband deployment can result from new entrepreneurship (Hasbi, 2020; Stephens et al., 2022). Indeed, faster NGA broadband can enable new firms to be formed which utilise data-intensive business models (not possible with slower broadband speeds), for example, by providing Software-as-a-Service (SaaS) products from the cloud to both businesses and consumers. Equally, demand for new data-intensive online content creates new market niches for businesses to serve (generally driven by the ubiquity of smartphones and other devices), driving business establishment growth.

The article is organised as follows. Section 2 summarises the relevant literature and Section 3 develops a theory of change that enables a testable hypothesis to be derived. The data we use is described in Section 4, and the econometric method employed to analyse them in Section 5. Section 6 presents the results of the econometric analysis, while Section 7 applies robustness tests to these results. Section 8 discusses the key findings, limitations and policy implications, before concluding.

---

[1] The MSOA is a very small geographical unit of analysis in the United Kingdom. The MSOA can be further sub-divided into Lower Super Output Areas (LSOAs). MSOAs represent subdivisions of the local authority districts (LADs), which are electoral administrative boundaries. LADs are then grouped into counties.

## 2. LITERATURE REVIEW

The general role of infrastructure (e.g., transportation) as an additional factor of production driving economic performance has been researched extensively in the last few decades (Aschauer, 1989). In more recent years the focus has been moving towards assessing the contribution of newer technologies from fixed telephony to Asymmetric-Digital-Subscriber-Line (ADSL) and FTTC broadband (Ford and Koutsky, 2005; Qiang et al., 2009), particularly given their role in driving the digital transformation of local areas and regions (Henderson, 2020; Haefner and Sternberg, 2020). Concerns have been expressed that inadequate provision of broadband infrastructure may severely disadvantage certain regions and firms, particularly in rural areas, leading to new digital inequalities (Riddlesden and Singleton, 2014; Oughton et al., 2015; Hilbert, 2016). Disparities in broadband often emerge due to market failure driven by (i) low demand-side adoption rates (Manlove and Whitacre, 2019; Kongaut and Bohlin, 2014; 2016), (ii) low supply-side economies of scale due to population density (Oughton et al., 2018a) and (iii) sub-optimal regulatory practices (Bauer, 2010; Cave et al., 2019).

How broadband affects economic performance can be analysed via several metrics, including economic growth and productivity (Whitacre et al., 2014; Arvin and Pradhan, 2014; Koutroumpis, 2009; Czernich et al., 2011; Tranos et al., 2020), income (Whitacre et al., 2014), employment (Jayakar and Park, 2013; Crandall et al., 2007; Kolko, 2012; Conley and Whitacre, 2016). Indeed, a number of studies reveal a positive impact of broadband adoption on economic growth at the national level (Holt and Jamison, 2009, Greenstein and McDevitt, 2011; Bertschek et al., 2013). Existing studies tend to focus on the contribution of broadband to economic growth in the US context and generally report positive relationships (e.g., Holt and Jamison, 2009; Whitacre et al., 2014). Additionally, positive relationships between broadband availability and employment are also found (Crandall et al., 2007). Koutroumpis (2009) investigates the effect of broadband across 22 OECD countries between 2002 and 2007, finding that broadband contributes up to 0.4% growth per year or 9.8% of total growth in gross domestic product during this period. An updated analysis reaffirms the previous conclusion (Koutroumpis, 2019). Similarly, Czernich et al. (2011) find that between 1996 and 2007, a 10% increase in broadband penetration led to 0.9-1.5% increase in the rate of per capita economic growth. Such positive effects are however heterogeneous, accruing mainly in areas with sufficient human capital resources and differ between industries, with manufacturing firms benefiting little compared to services (Haller and Lyons, 2015).

However, the impact of broadband on business creation is relatively less studied as a metric of economic performance. Although most studies report positive relationships between broadband and business creation, these appear to be nuanced (e.g., Audretsch et al., 2015; McCoy et al., 2018; Hasbi, 2020; Stephens et al., 2022; Luo, 2022). For example, several studies show that this relationship has heterogeneous effects across sectors (e.g., McCoy et al., 2018; Hasbi, 2020). Indeed, Mack and Grubesic (2009) conduct a study on the state of Ohio, using data from 1999 to 2004 with aggregate state results failing to demonstrate any significant relationship. Only when data are disaggregated by industrial sector does the relationship become observable. Smaller firms are also more likely to benefit from broadband compared to medium and large firms, and this relationship displays substantial spatial variations. More recently, Tranos and Mack (2016) investigate knowledge intensive firms and find bi-directional causality. Additionally, a study by McCoy et al. (2018) conducted in the Republic of Ireland report positive association between broadband and business creation, and that this effect becomes amplified in areas with high human capital, especially for high-tech firms. Hasbi (2020) examines the role of NGA broadband in France using micro-panel data, concluding that high speed broadband contributes to business creation within the tertiary and construction sectors, though the benefits are not evenly distributed and vary by area based on the endowment of human capital. Both studies find that the strongest relationship between broadband and business creation are exhibited in manufacturing, services (internationally traded), commerce and transportation.

There is also a growing body of rural-focused studies on the relationship between broadband and business activity (Kim and Orazem, 2016; Conroy and Low, 2021; Duvivier and Bussière, 2022; Stephens et al., 2022). Kim and Orazem (2016) examine the effect of broadband on new firm entry in rural areas in the US. Rural areas are typically regarded as disadvantaged regions with suboptimal infrastructure and deficient human capital. The study shows that the effect of broadband is strongest in those rural areas which border populous urban economies, as broadband facilitates connections to wealth-generating metropolitan areas. Conroy and Low (2021) evaluate the linkages between broadband and entrepreneurial activity in rural America between 2005 and 2007, finding a positive association. However, Duvivier and Bussière (2022) evaluate the impact of the French NGA broadband on business start-ups in rural municipalities between 2013 and 2018 and after allowing for spatial heterogeneity, report that positive broadband effects were limited to municipalities with favourable initial conditions in terms of local economic climate, natural amenities, and demographics.

Some scholars suggest that the reported positive economic effects of broadband on other variables besides firm creation at the local level may be over-stated. For example, while Kolko (2012) finds that broadband led to employment growth, there appears to be little, if any, effect on the local unemployment rate and average wage. This suggests that the apparent benefits of broadband may be through the relocation of factors of production from other areas, with the *net* benefit being more limited. Even if positive effects are achieved overall for a region, the non-divisibility of broadband may mean minimal benefits or even negative consequences for a particular local area. Ford and Seals (2019) argue that the evidence of job growth associated with broadband availability is limited at the local level. Indeed, faster speeds do not necessarily always demonstrate a significant economic payoff (Ford, 2018).

In this article we contribute to the growing literature by exploring the link between broadband and business creation at a very local level, in the hope of detecting some of the more nuanced relationships in this topic, with corresponding implications for policy. We focus explicitly on NGA broadband, which consists of a newer set of broadband infrastructure technologies, enabling speeds well in excess of those achieved via ADSL (from 24 Mbps to >200 Mbps). The much higher speed facilitates many data intensive applications (from Ultra High-Definition video streaming to intensive cloud usage) that offer opportunities for business creation, and which has generally received less attention in the literature.

3. THEORY OF CHANGE

In this section, a conceptual framework is defined. We consider the deployment of new broadband technologies as part of an ongoing cycle whereby new businesses are created, which in turn further increases demand for new broadband capacity. Hence, this is a co-evolving process where newly upgraded broadband infrastructure induces new economic opportunities in a long-term cyclical evolutionary process (Boschma and Martin, 2010; Sunley and Martin, 2020; Geels, 2002; Oughton et al., 2018b). Approximately every decade a major new fixed broadband technology is deployed, with past examples being the shift from dial-up Internet to ADSL, and then to NGA (as examined here in this article). In the 1990s downstream capacity was up to 8 Mbps, shifting to 24 Mbps with ADSL in the 2000s, and then more recently up to 200 Mbps or higher via NGA technologies. The applications that use Internet connectivity have a variety of preferred capacities for web browsing (0.33 Mbps), basic email (1 Mbps), streaming High Definition (HD) video (5 Mbps), cloud-based services (5 Mbps), file sharing (15 Mbps) and video calling (28 Mbps). Therefore, the available user capacity influences the types of content, applications and services which can be utilised (Stocker and Whalley, 2018). Generally, such higher capacity connections (often providing greater reliability) enable more types of online activities to be undertaken, improving information availability, increasing the endowment of human capital available in a local area. Good broadband access allows firms to use and sell sophisticated tools, such as cloud computing which can reduce capital expenditures and sunk costs (Duvivier et al., 2021). In this case, NGA broadband access provides many benefits for businesses, since sophisticated digital applications require higher connection speeds (Nicoletti et al., 2020), which legacy generation technologies are unable to provide (e.g., ADSL2+). Indeed, past studies have found that higher capacity broadband has a more beneficial impact on establishments than lower speed technologies. For example, recent research by Deller et al. (2022) shows that access to higher mobile speeds (25 Mbps) matters for rural entrepreneurship.

In the UK, access to broadband has been characterised by hesitant deployment and a lack of investment in the fibre infrastructure that enables higher broadband speeds. Many companies and households have thus been constrained in their access to higher broadband speeds, an impediment that has become more important with the advent of more sophisticated digital applications. Against this background, the availability of faster broadband in a local area improves its relative attractiveness as a place to start a business. The stock of human capital influences the ability for new business models and opportunities to be identified for selling goods and services. Activities which heavily rely on information transfer become more feasible, encouraging entrepreneurs to begin new business ventures across urban and rural areas, some of which are connected to customers primarily through digital means. A good example of this is the entire digital ecosystem developed around the availability and consumption of services enabled by broadband (Fransman, 2010). Although large businesses are generally involved in designing the hardware consumer devices used to access the Internet, thousands of small and medium-sized businesses develop content and services in the software application layer of the Internet. With new businesses being created, these firms have the potential to out-compete and destroy older, less efficient economic activities based on creative-destruction (Schumpeter, 2010). Consequently, the mass of new businesses created to take advantage of online service provision have a feedback effect on the aggregate demand for broadband capacity, thereby increasing the need for a new generational broadband upgrade.

Given this theory of change, we hypothesise that the net effect of NGA broadband on business creation at the local level is likely to be positive, as is supported within the existing broadband literature (Koutroumpis, 2009; Czernich et al., 2011; Tranos et al., 2020; Whitacre et al., 2014; Jayakar and Park, 2013; Hasbi, 2020). Moreover, the evolutionary cycle culminating in creative-destruction means that the effect of broadband are heterogenous across different types of local areas. We also explore some of these nuances. In the next section we outline the data that we employ to test this hypothesis on the relationship between NGA broadband provision and business creation.

4. DATA

We utilise annual data at the MSOA level in England for 2011 to 2015, which is among the highest geographical granularity when compared to existing studies. MSOAs are sub-divisions of Local Authority Districts (LAD) in England, which functions as electoral boundaries. England has an area roughly 130,000 km$^2$ in size and a population of 54.3 million in 2015. This area is divided into 6,791 MSOA units, which means that an average MSOA is only 19 km$^2$ and has a population of about 8,000. Table 1 below lists two studies which are comparable to ours in terms of the spatial resolution employed. As can be seen, our level of resolution is considerably higher than McCoy et al. (2018), and comparable to the recent study by Hasbi (2020), which is the most granular study we find, though the study focuses on urban areas outside of the biggest cities of Paris, Lyon and Marseilles. McCoy et al. (2018) had to make a trade-off between using very fine geographical data, and too many local statistical areas containing no business establishments (the "excess zero problem"). The trade-off is necessary in the case of Republic of Ireland, a relatively sparsely populated country. As England is more densely populated, all MSOA areas contain business establishments hence the trade-off question does not apply in our case, and there is no need for us to employ more aggregate levels of geography (e.g., to the LAD or county levels).

For a complete list of data and associated descriptions, see Table A1 in the Appendix, as well as Table A2 containing descriptive statistics. Our dataset covers the years from 2011 to 2015, across 6,791 area units (MSOAs) in England each year.

Table 1. Evaluating broadband and business creation using high resolution data

| Study | Hasbi (2020) | McCoy et al. (2018) | Mack et al. (2011) |
|---|---|---|---|

| Country of analysis | France[2] | Republic of Ireland[3] | Six metropolitan areas in the United States |
|---|---|---|---|
| Year | 2010-2015 | 2001-2011 | 2001-2006 |
| Area under study | Approximately 90,000km$^2$ of France's total area of 643,801km$^2$ | Approximately 16,547km$^2$ of Republic of Ireland's total area of 70,273km$^2$ | Atlanta, Boston, Columbus, Dallas, Detroit and San Jose, total area: 10,578km$^2$ |
| Population under study | Approximately 54 million of France's 67.5 million | 3.2 million of the Republic of Ireland's 4.5 million | Approximately 4.7 million of the USA's 320 million |
| Area unit number | 4,935 | 858 aggregated to 192 | 410 |
| Area unit classification | Municipalities or arrondissements | Electoral division aggregated into "urban fields" by authors' calculation | US ZIP codes |
| Average size per area unit | Approximately 18km$^2$ | 86km$^2$ | 25.8km$^2$ |
| Average population per unit area | Approximately 12,000 | 16,893 | 11,420 |

*4.1. Dependent variable*

The dependent variable in our economic model is the annual rate of change in the number of business establishments for the local statistical areas, which is calculated using business counts from the UK Office of National Statistics (ONS) Nomis (2018). Figure 1 shows geographically the net percentage growth in business establishments during the study period. It should be noted that our data denotes *net* percentage change in the number of business establishments. We use this since the data for establishment entry, exit and survival are not available in England at the MSOA level. Moreover, using net change has the advantage of reflecting the area's ability to both attract new entrants, as well as to discourage existing businesses from relocating or cease operating.

---

[2] Only urban areas or municipalities are examined, covering approximately 75% of the population, with the largest urban centres of Paris, Lyon and Marseille excluded from the analysis.
[3] Only settlements which are above the 75th percentile in terms of population density were included as McCoy et al. (2018) argue that other areas are not natural locations for businesses and their inclusion would lead to the "excess zero problem" whereby a large number of local units contain no business establishments. We replicate their data to estimate the area size and population this study actually included in the analysis.

Figure 1: Net percentage growth in business establishments over the study period (2011-2015)

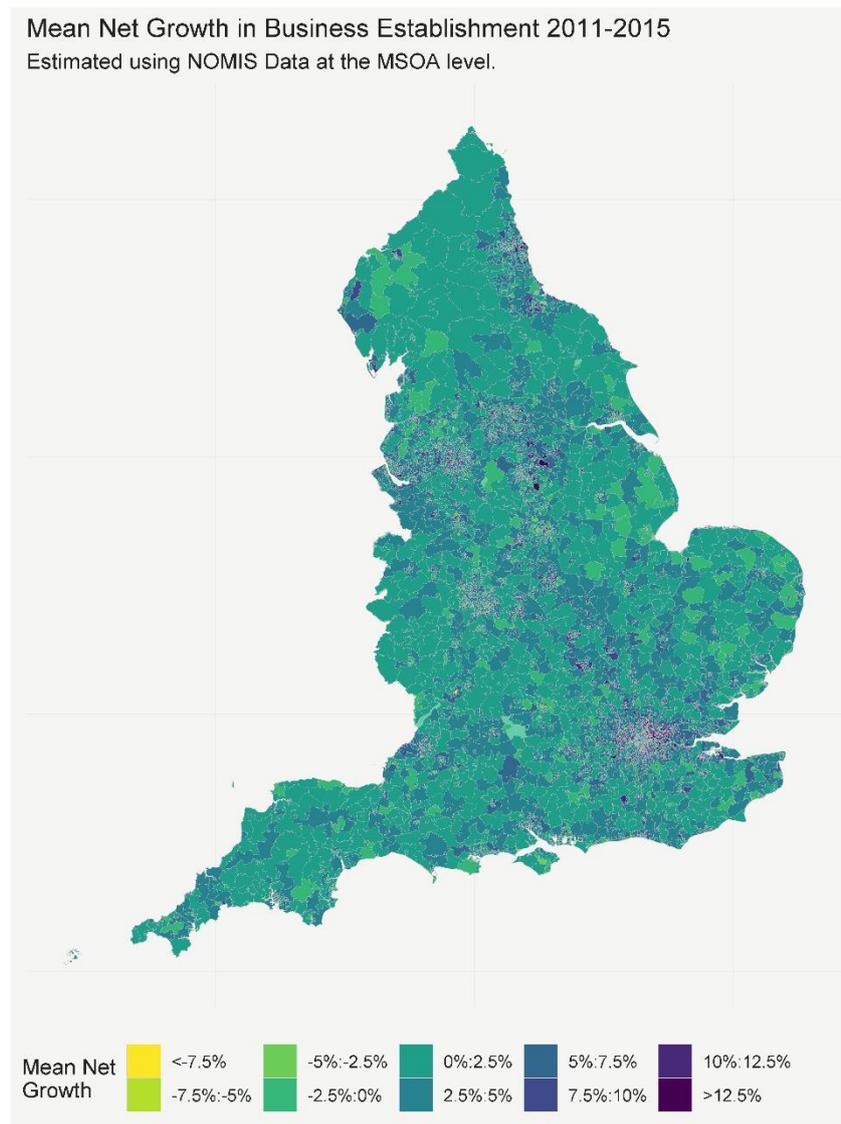

*4.2. Explanatory broadband variable*

We utilise download speed as the explanatory (independent) variable of interest. Thus, a crowdsourced Broadband Speed Checker dataset was obtained for 2011-2015 (n=11,052,701) representing the real broadband speed experienced by end-users (Broadband Speed Checker, 2021). Speed Checker is one of the best tools to represent realistic broadband experience, without needing to place a physical device in each user's premises to collect data which can be difficult to design and administer. These 11 million individual data points on recorded broadband speeds were mapped to MSOA local areas using geographical information system (GIS) software and ONS lookup layers. These data points were then used to calculate the download speed in each MSOA region in each of the years between 2011 and 2015. Figure 2 shows the download speed across MSOAs in England as averaged over the study period.

Figure 2. Average download speeds over the study period (2011-2015)

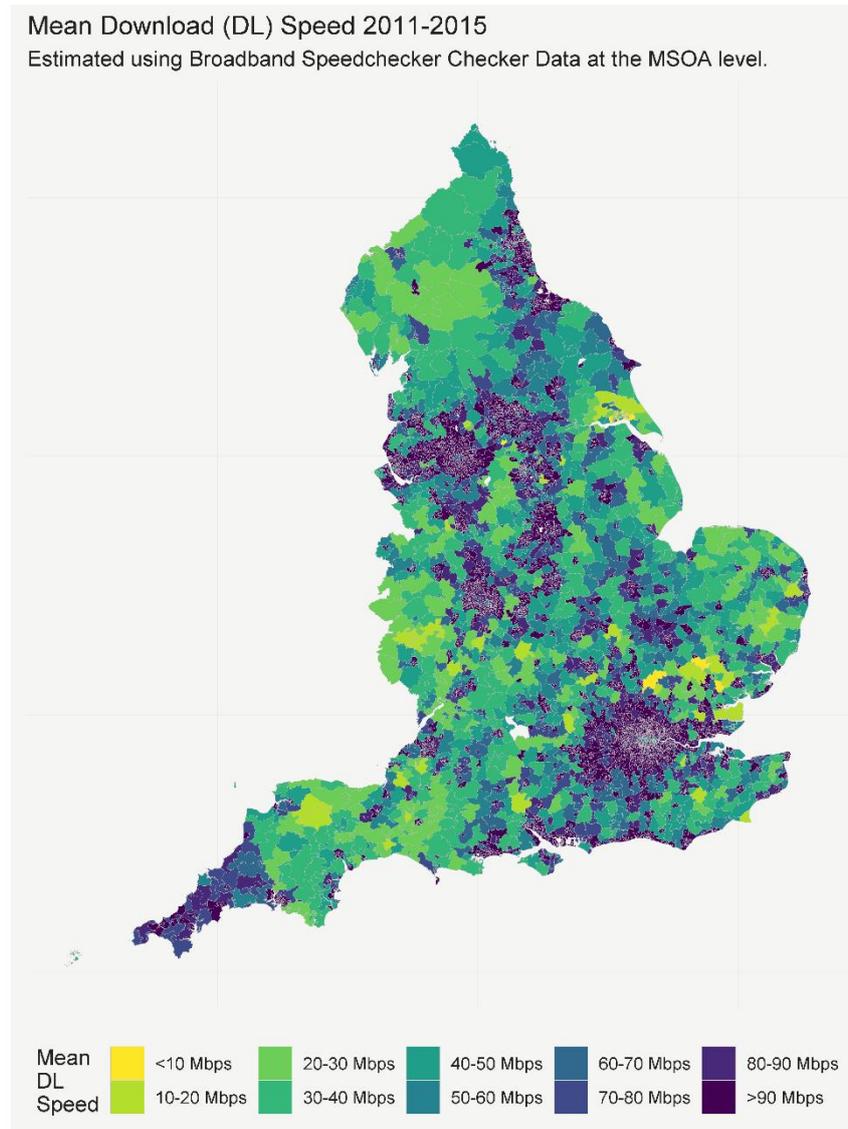

*4.3. Control variables*

The control variables we include can be divided into several groups. First, we include local area population to proxy for scale effect, as larger areas with a higher base number of establishments are less likely to maintain a high *rate* of growth ceteris paribus, than one with a small base number of establishments.[4] Second, we include a group of variables that reflect the quality of local amenities and living conditions that are likely to influence prospective employees' decisions to locate to the area to live and work and be hired. These include road density – a proxy for the quality of local transport infrastructure and connectivity; inverse travel time to nearest town centre – a proxy for access to local services; inverse travel time to (nearest) employment centre – reflects convenience to work or potential work; inverse travel time to (nearest) supermarket – reflects the quality of local amenities and convenience of living; and lastly, crime rate – an indicator for local area deprivation and attractiveness to live. Third, we include a group of variables to reflect local economic and market conditions that are likely to attract prospective firms to create an establishment in the area. These include labour cost – a proxy for attractiveness and availability of labour force; the Herfindahl-Hirschman (HH) index that reflects sectoral concentration – proxying for the degree of local business competition; population density – which reflects the degree of urban concentration; and finally, the unemployment rate – capturing the level of labour market slackness. These control variables are selected based on those that have been commonly used in past studies (such as Kolko, 2012; Mack et al., 2011; Mack and Rey, 2014). Below we describe briefly how they are compiled.

Population is easily obtained from the Office for National Statistics (ONS). The inverse travel times to town centres, supermarkets and employment centres are obtained from the UK Department for Transport accessibility index which is based on the travel time by foot or public transport (DfT, 2011; 2018). Crime rate and antisocial behaviour data are obtained from the UK police websites (Data.Police, 2020). The labour cost index is constructed using the ONS quarterly estimates of labour cost divided by industrial sectors, seasonally adjusted. We average the quarterly estimates to produce annual figures. We then use the proportion of employment in each sector for local statistical areas to compute a weighted index of the labour cost. The division of labour cost and employment proportion by sector are based on the broad industrial group list.

---

[4] This is similar to McCoy et al. (2018) including total employment as a scale variable. Population and employment are always extremely highly correlated, yet we choose population overall as we have another control variable, inverse travel time to nearest employment centre, which may be more correlated with total employment than with population size, hence the latter is slightly more preferrable.

For the HH index of sectoral employment concentration, we calculate using the following formula:

$$HH_i = \sum_{j=1}^{J} s_{ij}^2$$

Where $j = 1,2,...J$ denotes each industrial sector, while $s_{ijt}$ refers to the share of employment of sector $j$ in area $i$. The division of employment by sector is also based on ONS broad industrial groups.

Data for the population density and unemployment rate are also obtained from the ONS.

*4.4. Instrumental variable (IV)*

We rely on an instrumental variable approach (IV) to deal with potential endogeneity between broadband speed and the growth rate of business establishments. We select an IV that is a strong predictor of broadband download speeds but is unlikely to have had a direct impact on the business establishment growth rate during the studied period. Specifically, we use ADSL broadband coverage from 2001-2005 as the instrument for download speeds in years 2011-2015. Each year's data for download speed is instrumented by the ADSL data ten years prior, so 2011 is instrumented by 2001, whereas 2012 is instrumented by 2002 etc. ADSL broadband was the previous generation of broadband technology before the deployment of NGA FTTC. ADSL broadband coverage averaged in 2001 is shown in Figure 3. The coverage of ADSL broadband is a derived variable. Telephone exchange locations and serving postcodes are identified using a postcode dataset available from SamKnows (2020) to obtain the number of premises per postcode. This includes an allocated ADSL-enabled date depending on when the telephone exchange was upgraded. The ADSL coverage across England had taken place rapidly between 2001 and 2005 across many local areas, while the majority of NGA broadband deployment had taken place between 2011 and 2015, around a decade apart.

The selection of ADSL as an Instrument follows the approaches in Duranton and Turner (2012) and Möller and Zierer (2018), where the authors employ measures of legacy infrastructure as IVs. An important potential source of endogeneity stems from any unaccounted-for contemporary events that could have affected the download speeds and business creation simultaneously, resulting in an omitted variable bias. For example, an improvement in the quality of local authorities' governance capabilities may positively affect both variables. As such events do not affect past patterns of infrastructure roll-out, using past ADSL coverage data helps to eliminate this potential source of endogeneity.

However, the persistence of local conditions can affect business creation long into the future, which would contribute to the bias of our IV. For example, favourable demand conditions may have driven ADSL roll-out in 2001-2005, with the same conditions continuing to affect business creation and download speed in 2011-2015. Nonetheless, the literature on 'productivity paradox' (e.g., Crafts, 2018) implies that the economic conditions in the early 2000s and 2010s are not comparable, with the former period being associated with rapid productivity growth via digital technologies, while the later with declining growth despite heavy digital investments. The change in economic conditions between the two periods weakens the persistence of factors that influenced ADSL roll-out in 2001-2005, as these factors are less likely to continue affecting business creation in 2011-2015 due to the change. This source of bias can be further reduced by adopting a panel data fixed-effects model, which uses model transformation to remove time-invariant heterogeneity at the local level. We discuss this modelling approach in the next section.

Figure 3. Instrumental variable: Percentage of premises with ADSL broadband access (2001)

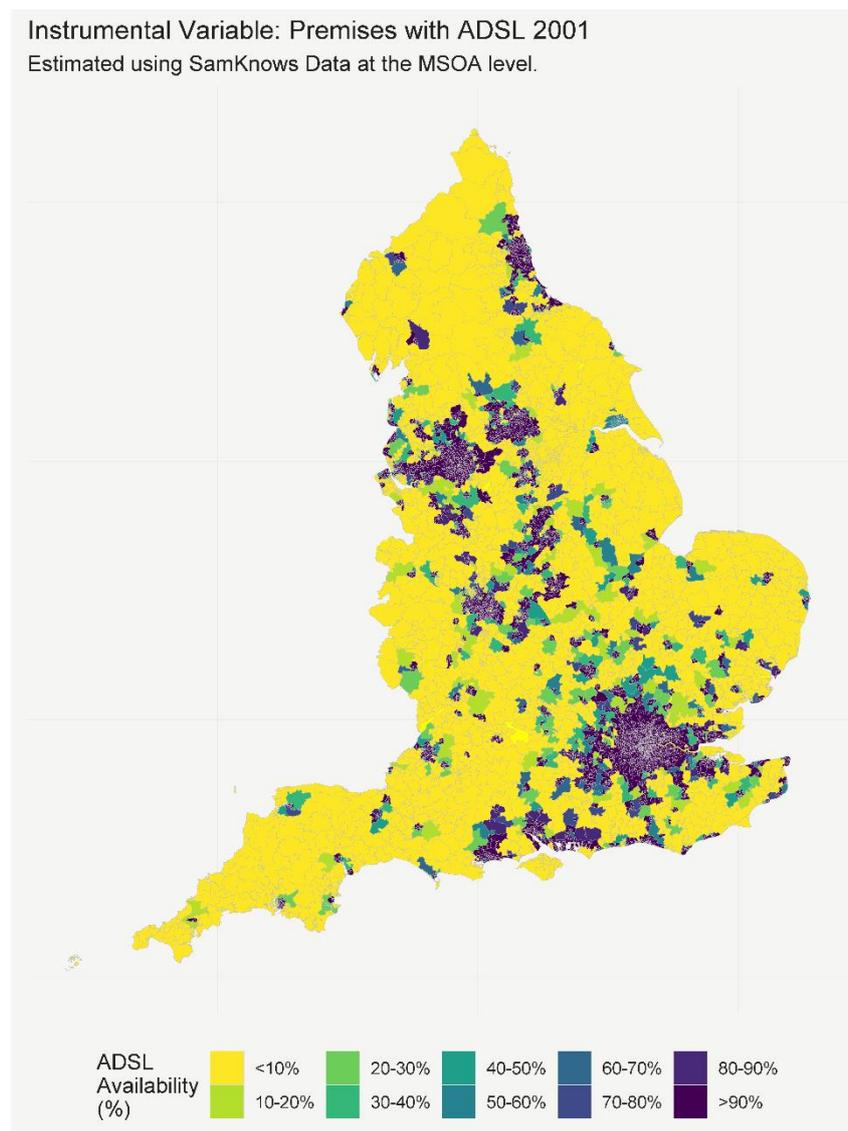

## 5. METHODOLOGY

*5.1. Econometric modelling*

We employ a panel data model, which takes on the following form:

$$(1) \quad \%change\_estab_{it} = \alpha_0 + \alpha_i + \beta \ln(download\_speed_{it}) + \gamma X_{it} + \theta \%change\_estab_{it-1} + \varepsilon_{it}$$

Where $\%change\_estab_{it}$ denotes the percentage change (rate of change) in the number of business establishments in area $i$ in year $t$; $download\_speed_{it}$ is the indicator of the broadband download speed in area $i$ at time $t$, expressed in natural logs, with $\beta$ as the coefficient to be estimated; $X_{it}$ a vector of control variables with a vector of corresponding coefficients denoted by $\gamma$; we include the lagged value of the dependent variable $\%change\_estab_{it-1}$ to account for serial correlation, which is captured by the coefficient $\theta$; $\alpha_0$, $\alpha_i$ and $\varepsilon_{it}$ represent the intercept term, local area time invariant heterogeneity and the error term, respectively.

Some of the control variables included in $X_{it}$ are converted into natural logarithm forms. This is so that their coefficients can be interpreted as a relationship of elasticity (i.e., percentage change versus percentage change) against the dependent variable. Not all variables in $X_{it}$ however are suitable for this conversion since some are naturally expressed as a percentage value. For example, the control variable unemployment rate has a range of 0 and 100%, and it would not be appropriate to apply a natural log transformation.[5] For an indication of which variables are logged and which remain unchanged, please refer to Table A2 in the Appendix, which describes our data.

We estimate model (1) using the fixed-effects estimator. The fixed-effects model is strongly preferred to the random-effects model as it eliminates the effect of time invariant heterogeneity, or $\alpha_i$ in our model (Wooldridge, 2010). As the independent variable of download speed is in natural log form, the coefficient $\beta$ should be interpreted as an elasticity.

---

[5] If natural logs are taken for these variables, then they would become a percentage change upon a percentage change, which would be difficult to interpret.

*5.2. Addressing endogeneity*

In the presence of endogeneity in equation (1), the standard single-stage model may result in biased estimates of $\beta$, which do not adequately identify the causal relationship of download speed on business creation. To deal with the potential endogeneity, we apply a Two-Stage-Least-Squared (2SLS) model, using an instrumental variable (IV) to exogenise the variable for download speed. The first stage equation is specified as equation (2):

$$(2) \quad \ln(download\_speed_{it}) = \theta\%change\_estab_{it-1} + \alpha_0 + \alpha_i + \mu \ln(IV_{it-j}) + \gamma X_{it} + \varepsilon_{it}$$

Whilst $\%change\_estab_{it-1}$ accounts for serial correlation in equation (1), its main function in equation (2) is to reduce potential endogeneity (from reverse causality), controlling for broadband deployment at time $t$ being affected by expectations of past business growth rates. This approach is similar to Hasbi (2020). $IV_{it-j}$ refers to the instrumental variable in area $i$ at time $t-j$, where $j = 10$. Its coefficient on the endogenous variable download speed is shown by $\mu$. This method is based on the approach of Duranton and Turner (2012), who use the initial values of existing patterns of infrastructure deployment in the distant past as an instrument for current stock of infrastructure (independent variable).

Importantly, our vector of control variables $X$ includes various proxies of demand conditions for broadband, such as population density and distances to town centres, supermarkets, and employment hubs. Therefore, our first stage controls for and alleviates endogeneity arising from persistently favourable demand conditions. A similar instrumental variable approach has been employed by Czernich et al. (2011), Bertschek and Niebel (2016) and Stockinger (2019).

The second stage equation is specified by equation (3):

$$\text{(3)} \quad \%change\_estab_{it} = \alpha_0 + \alpha_i + \hat{\beta}\ln(\widehat{download\_speed}_{it}) + \gamma X_{it} + \theta\,\%change\_estab_{it-1} + \varepsilon_{it}$$

Where $\widehat{download\_speed}_{it}$ denotes the fitted values of download speed using the parameters estimated in the first stage equation. In the second stage equation coefficient $\hat{\beta}$ identifies the exogenous relationship between establishment growth rate and download speed.

*5.3. Data stratification*

To explore some of the nuances in the relationship between broadband and business creation, we re-run the regression in equations (1) and (3) stratifying the total number of statistical areas (6,791) into various subsets to examine the differential effects across localities that have different characteristics. We stratify MSOA areas based on London/non-London local statistical regions, as we suspect that as an exceptionally major metropolitan area London could be differently affected by NGA broadband adoption, compared to the rest of England. A key reason for this might be high reliance on dedicated fibre optic connections provided to large corporate or residential buildings. Also, the patterns of broadband technology deployment could be different in London. Hence, we re-run the models separately for localities within the London region, and for the rest of England. This approach is similar to that used by McCoy et al. (2018), who exclude Dublin in their analysis of the Republic of Ireland. We also stratify areas by different levels of human capital, separating them into areas with the proportion of workers with tertiary education less than 35%, between 35 and 45% and above 45%, which we label "low", "medium" and "high" human capital areas, respectively. This numeric division is selected as it separates the total number of local statistical areas in England into three groups roughly equal in number. Stratification by human capital is of interest as, hypothetically, broadband speeds could have differential effects depending on types of businesses dominating the local economies, with the growth in the more high-tech business areas (associated with areas higher in human capital) being more dependent on new technologies such as NGA broadband. Table 2 below presents the average annual change in business establishments separated by the stratified areas.

Table 2. Average annual growth rate in the number of businesses (2011-2015), by sub-sample

| Metric | All statistical areas | London | Outside London | Low Human Capital | Medium Human Capital | High Human Capital |
|---|---|---|---|---|---|---|
| Mean | 4.03 | 7.37 | 3.47 | 4.16 | 3.89 | 4.01 |
| Sta deviation | 3.55 | 4.08 | 3.12 | 4.01 | 3.29 | 3.13 |
| N | 6,791 | 983 | 5,808 | 2,681 | 2,140 | 1,970 |

6. RESULTS

The estimates for our single-stage regression models, i.e. the estimation of equation (1), are presented in Table 3. The estimated coefficient on broadband speed is positive and statistically significantly associated with the percentage change in business establishments across all six models. Model (1) is the default estimate which employs the complete set of data; Models (2) and (3) are regression models stratified using sub-samples of local areas located within London and outside, respectively; Models (4), (5), and (6) show the results using sub-samples of local areas in the low, medium and high human capital level groups, respectively. Standard errors are reported in parentheses and all our models apply cluster-robust standard errors. The corresponding 2SLS model estimates are shown in Table 4.

Although the results in Table 3 establish a starting point, we suspect that the use of single-stage estimation would be affected by endogeneity. Hence, the results in Table 3 serves as a point of departure and comparison with our 2SLS estimates presented in Table 4. For Model (1) in Table 4, where we use the complete set of data to run the regression, our result obtained using 2SLS shows a positive and statistically significant impact of download speeds on business establishment growth, with a one percentage increase in broadband speeds being associated with 0.0574 percentage points faster annual growth in the number of business establishments. The estimated impact of broadband is similar in magnitude to that reported by Falk and Hagsten (2021), who find that this figure is between 0.023 and 0.050 in Sweden. It is also similar to, albeit somewhat higher than the estimate of Hasbi (2020), who reports 2.7% higher annual growth in business establishments for areas where NGA broadband was available. However, both of these studies are not directly comparable due to the measurement for broadband not being based on average speeds but on the accessibility of broadband that exceeds a certain speed (at least 30 Mbps for Hasbi, and at least 100 Mbps for Falk and Hagsten).

For London-based MSOAs, model (2) in Table 4 shows a negative and insignificant relationship. However, we show in further sections that this result is not reliable because historical ADSL coverage is a relatively weak instrument for London's MSOAs. For the rest of England as shown in model (3) of Table 4, the coefficient is slightly higher than in model (1) at 0.0617. Models (4), (5), and (6) in Table 4 show the results for sub-samples with low, medium and high levels of human capital, respectively. The coefficient for low human capital areas is the highest, at 0.0664. Medium human capital level areas have the weakest relationship between broadband and business establishment growth, with a one percentage increase in download speeds being associated with a 0.0448 percentage increase in the growth of business establishments. For areas classed as high-level human capital, this effect is estimated at 0.0473.

The signs of the coefficients on the control variables in all the models in Tables 3 and 4 generally conform to expectations. There is a negative association between the growth rate of business establishments and (log) of population, indicating that larger areas tend to have a smaller *rate* of growth. Road density does not appear to have a statistically significant effect on business creation except in areas with medium levels of human capital, and only at the 5% level of significance; Inverse time to (nearest) town centre generally shows negative associations with business creation, indicating that the growth rate of business establishments tend to be outside of the central business districts during the study period; Inverse travel time to (nearest) employment centre shows positive and highly significant association with business creation, indicating that the fastest growth occurred close to these employment centres, as is naturally expected; Inverse travel time to a supermarket does not show statistically significant relationship with business creation except in medium human capital areas and only at 10% level; crime is negatively and statically associated with business creation in most cases as would be expected, and for London regions in particular the negative coefficient is considerably larger than for all of England (-0.138 versus -0.0166). This shows that business creation is more sensitive to the crime rate in London, which is plausible given its heavily metropolitan nature; labour cost also shows negative association with business creation as would be expected; The HH index generally shows negative association which is to be expected given that more employment concentration may signal greater barriers for new firms to enter; population density is not statistically significantly associated with business creation except in areas of medium human capital levels; and lastly the unemployment rate is negatively and significantly associated with business creation.

Table 3. Regression results of single-stage panel data estimation. Dependent variable is the annual growth rate in the number of business establishments. Sample period: 2011-2015

| Variable/Model | (1) Entire Sample | (2) London | (3) Outside London | (4) Low Human Capital | (5) Medium Human Capital | (6) High Human Capital |
|---|---|---|---|---|---|---|
| Download speed | 0.0376*** (0.00238) | 0.0248*** (0.00622) | 0.0388*** (0.00244) | 0.0300*** (0.00514) | 0.0329*** (0.00411) | 0.0395*** (0.00427) |
| Population | -0.379*** (0.0487) | -0.128 (0.152) | -0.421*** (0.0494) | -0.135 (0.0901) | -0.262*** (0.0907) | -0.337*** (0.0553) |
| Road density | 0.0163 (0.0138) | 0.00641 (0.0200) | 0.0153 (0.0151) | 0.0290 (0.0220) | 0.0235** (0.0102) | -0.0123 (0.0251) |
| Inverse time to town centre | -0.00569 (0.00528) | -0.00937 (0.0169) | -0.00323 (0.00566) | -0.00876 (0.0109) | -0.00283 (0.00622) | 0.00141 (0.00820) |
| Inverse time to employment hub | 0.0258*** (0.00244) | 0.00542 (0.00733) | 0.0234*** (0.00262) | 0.0263*** (0.00669) | 0.0192*** (0.00305) | 0.0124*** (0.00373) |
| Inverse time to supermarket | -0.00702 (0.00614) | -0.00512 (0.00914) | -0.00829 (0.00619) | 0.00850 (0.0114) | 0.00600 (0.00633) | -0.000687 (0.00704) |
| Crime rate | -0.0175** (0.00798) | -0.0924*** (0.0212) | -0.00561 (0.00877) | -0.00932 (0.0169) | -0.0247*** (0.00782) | -0.0371*** (0.00823) |
| Labour cost | -0.960*** (0.121) | -1.730*** (0.225) | -0.898*** (0.106) | -0.854*** (0.111) | -1.224*** (0.101) | -1.510*** (0.138) |
| HH Index | -0.113** (0.0440) | -0.0201 (0.0676) | -0.120** (0.0512) | -0.0573 (0.0737) | -0.131* (0.0750) | -0.152** (0.0667) |
| Population density | 0.0803 (0.0734) | -0.0859 (0.137) | 0.0720 (0.0641) | 0.146 (0.100) | 0.255*** (0.0838) | 0.0349 (0.105) |
| Unemployment | -2.186*** (0.540) | -2.249*** (0.515) | -2.198*** (0.546) | -2.982*** (0.715) | -1.448** (0.602) | 0.818* (0.451) |
| % change in estab, lagged 1 | -0.177 (0.118) | 0.241* (0.126) | -0.223** (0.104) | -0.270*** (0.0958) | 0.00143 (0.0297) | 0.273*** (0.100) |
| Constant | 9.264*** (0.771) | 12.56*** (1.577) | 9.258*** (0.680) | 6.334*** (0.898) | 9.545*** (1.079) | 12.27*** (1.016) |
| N | 27,164 | 3,932 | 23,232 | 10,724 | 8,560 | 7,880 |

*, ** and *** denote statistical significance at the 10%, 5% and 1% level, respectively

Table 4. Regression results of 2SLS panel data estimation. Dependent variable is the annual growth rate in the number of business establishments. Sample period: 2011-2015.

| Variable/Model | (1) 2SLS Entire Sample | (2) 2SLS London | (3) 2SLS Outside London | (4) 2SLS Low Human Capital | (5) 2SLS Medium Human Capital | (6) 2SLS High Human Capital |
|---|---|---|---|---|---|---|
| Download speed | 0.0574*** (0.00471) | -0.0767 (0.107) | 0.0617*** (0.00491) | 0.0664*** (0.0139) | 0.0448*** (0.00674) | 0.0473*** (0.00609) |
| Population | -0.396*** (0.0480) | 0.0000572 (0.156) | -0.438*** (0.0494) | -0.159* (0.0888) | -0.265*** (0.0890) | -0.341*** (0.0548) |
| Road density | 0.0112 (0.0134) | 0.0134 (0.0221) | 0.00894 (0.0145) | 0.0240 (0.0210) | 0.0207** (0.0103) | -0.0150 (0.0245) |
| Inverse time to town centre | -0.0119** (0.00554) | -0.0356** (0.0174) | -0.0116* (0.00600) | -0.0176 (0.0115) | -0.00722 (0.00655) | -0.000648 (0.00834) |
| Inverse time to employment hub | 0.0254*** (0.00238) | 0.0117 (0.00785) | 0.0230*** (0.00263) | 0.0243*** (0.00664) | 0.0196*** (0.00306) | 0.0124*** (0.00377) |
| Inverse time to supermarket | 0.00311 (0.00682) | -0.00290 (0.0102) | 0.00580 (0.00702) | 0.0151 (0.0119) | 0.0144* (0.00748) | 0.00368 (0.00781) |
| Crime rate | -0.0166** (0.00797) | -0.138*** (0.0268) | -0.00543 (0.00874) | -0.00749 (0.0168) | -0.0245*** (0.00783) | -0.0364*** (0.00831) |
| Labour cost | -1.026*** (0.122) | -1.615*** (0.238) | -0.977*** (0.107) | -0.883*** (0.113) | -1.263*** (0.103) | -1.560*** (0.145) |
| HH Index | -0.103** (0.0440) | -0.0277 (0.0707) | -0.108** (0.0512) | -0.0469 (0.0738) | -0.122* (0.0739) | -0.148** (0.0665) |
| Population density | 0.0413 | -0.0389 | 0.0168 | 0.102 | 0.237*** | 0.0206 |
| | (0.0727) | (0.140) | (0.0644) | (0.100) | (0.0815) | (0.105) |
| Unemployment | -1.686*** | -4.457*** | -1.617*** | -2.142** | -1.006 | 1.012** |
| | (0.559) | (0.804) | (0.566) | (0.844) | (0.626) | (0.474) |
| % change in estab, lagged 1 | -0.181 | 0.253** | -0.228** | -0.270*** | -0.00510 | 0.264** |
| | (0.115) | (0.122) | (0.101) | (0.0944) | (0.0297) | (0.104) |
| Constant | 9.850*** | 10.93*** | 9.931*** | 6.696*** | 9.811*** | 12.62*** |
| | (0.781) | (1.607) | (0.698) | (0.919) | (1.075) | (1.030) |
| N | 27,164 | 3,932 | 23,232 | 10,724 | 8,560 | 7,880 |

*, ** and *** denote statistical significance at the 10%, 5% and 1% level, respectively

# 7. ROBUSTNESS CHECKS

We perform several types of robustness checks on our results to increase confidence in their validity. The first type of robustness checks examines the sensitivity of our results to different model specifications. The second type tests for the validity and strength of the chosen instrument. The third type examines the sensitivity of our results to sectoral classification of firms. Lastly, we check whether the coefficients for download speed in our stratified dataset for low, medium and high human capital areas are significantly different to each other.

*7.1. Alternative model specifications*

For robustness checks involving alternative model specifications, the results are shown below in Table 5. In model (7), we restrict the estimation sample to include only the localities which had a sufficient number of individual crowdsourced download speed measurements to minimise the variance and error in the average download speeds. While we rely on millions of individual observations of download speeds to compute the MSOA averages, for some areas, the number of data points is relatively small, resulting in larger variance of the estimated MSOA averages. Hence, we test the robustness of the results by limiting the estimation sample to MSOAs where the coefficient of variation of the computed average download speeds does not exceed 20%. We re-run the 2SLS model with these observations removed, leaving us with an unbalanced panel of 24,220 observations. The results remain statistically significant and the magnitude little changed, at 0.0600, as shown in model (7) in Table 5.

We also re-run the 2SLS model with all the variables in level forms instead of natural logarithm, as shown in model (8). Once again, download speed is positively and significantly associated with business creation. The coefficient however needs to be interpreted differently – an increase in download speed by 1 Mbps is associated with an increase in the growth rate of business establishment by 0.00551 percentage points. We also include the lagged values of the download speed, since it is possible that broadband speeds can affect business growth with a lag. We find that in this model, the contemporaneous effect of broadband is reduced compared to the main 2SLS model, with the coefficient being at 0.0348 though still significant, as shown in model (9). The coefficient for the lag of download speed is 0.0260 and is significant. The sum of these two coefficients, 0.0608 is similar to that of the main 2SLS model at 0.0574, which suggests that the effect of download speed is split between the contemporaneous year and the previous year.

To test whether our result is sensitive to the effects of spatial dependencies, we apply a spatial Durbin model. This involves computing a spatially-weighted measure of broadband speeds in nearby areas in order to control for the effect of broadband infrastructure on neighbouring areas. The most common spatial weight matrices for geographical proximity are the contiguity and distant decay matrix. We employ the former, which is significantly less computationally demanding. Given that our purpose is only to control for, rather than to produce, a quantitatively accurate estimate of spatial dependence, this simpler method suffices. For contiguity matrices, geographically adjacent areas are designated with a value of 1, while all others are assigned 0. The contiguity matrix is then multiplied by broadband download speed to generate the variable of weighted neighbouring broadband infrastructure, which is included into the regression as an additional variable. As seen in model (10), the coefficient from the spatial Durbin model is 0.0147, which is substantially reduced in magnitude compared to models that do not include the spatially-weighted measures of broadband speed. The spatially-weighted download speed displays a positive and significant coefficient at 0.0441. The sum of the two coefficients at 0.0588 is very similar to the main coefficient of 0.0574. This suggests that there is a sizable positive spill-over effect of broadband speeds from neighbouring areas that enhances business creation in a local area. This is contrary to many studies which report a negative displacement effect of infrastructure. Our finding of a positive relationship may be due to the analysis being conducted at a finer geographical level, where many of the benefits from broadband would be indivisible. Instead, a local area network effect spanning multiple MSOAs jointly benefit from broadband. The reduction in the coefficient of download speed when the spatial Durbin is applied suggests that the quality of broadband in neighbouring areas partially explains business creation. However, even accounting for this effect, the relationship remains positive and statistically significant. We also apply the Moran's I test to the residuals in our default 2SLS regression model and that of the spatial Durbin model, shown below in Table 6. The Moran's I results are not statistically significant, and the application of the spatial Durbin model only marginally reduces the Moran's I, indicating that our estimates of the coefficients are unlikely to suffer from bias arising due to spatial dependencies.

Table 5. Robustness-checks with alternative model specifications

| Variable/Model | (7) Excess variance in download speed removed | (8) Level form model | (9) Lagged download speed model | (10) Spatial Durbin model |
|---|---|---|---|---|
| | 0.0600*** | 0.00551*** | 0.0348*** | 0.0147*** |
| Download speed | (0.00517) | (0.000444) | (0.00647) | (0.00236) |
| Population | -0.424*** | -0.0000334*** | -0.403*** | -0.401*** |
| | (0.0513) | (0.00000706) | (0.0474) | (0.0478) |
| Road density | 0.0126 | 1.735 | 0.00749 | 0.00986 |
| | (0.0149) | (2.094) | (0.0133) | (0.0135) |
| Inv. time to town centre | -0.00864 | -0.118 | -0.0113** | -0.0138** |
| | (0.00588) | (0.0807) | (0.00546) | (0.00538) |
| Inv. time to employ hub | 0.0237*** | 0.160*** | 0.0222*** | 0.0257*** |
| | (0.00242) | (0.0193) | (0.00220) | (0.00233) |
| Inv. time to supermarket | 0.00204 | -0.00187 | 0.00318 | 0.00267 |
| | (0.00745) | (0.0405) | (0.00677) | (0.00602) |
| Crime rate | -0.00983 | -0.000776 | -0.0194** | -0.0144* |
| | (0.00901) | (0.000685) | (0.00756) | (0.00808) |
| Labour cost | -1.027*** | -0.00252*** | -1.090*** | -1.023*** |
| | (0.127) | (0.000250) | (0.111) | (0.116) |
| HH Index | -0.0975* | -0.0915** | -0.0973** | -0.102** |
| | (0.0523) | (0.0446) | (0.0443) | (0.0440) |
| Population density | 0.0111 | -0.000757 | 0.0150 | 0.0330 |
| | (0.0824) | (0.00137) | (0.0731) | (0.0727) |
| Unemployment | -1.368** | -0.917 | -1.392*** | -1.565*** |
| | (0.665) | (0.560) | (0.507) | (0.504) |
| Download speed lagged | - | - | 0.0260*** | - |
| | | | (0.00485) | |
| Weighted download speed | - | - | - | 0.0441*** |
| | | | | (0.00410) |
| % change in estab, lagged 1 | -0.191 | -0.185 | -0.189* | -0.183 |
| | (0.121) | (0.113) | (0.114) | (0.116) |
| Constant | 10.18*** | 1.512*** | 10.35*** | 9.799*** |
| | (0.771) | (0.130) | (0.716) | (0.737) |
| N | 24,220 | 27,164 | 27,164 | 27,164 |

*, ** and *** denote statistical significance at the 10%, 5% and 1% level, respectively

Table 6. Moran's I on model residuals, 2011-2015

| Year | 2011 | 2012 | 2013 | 2014 | 2015 |
|---|---|---|---|---|---|
| 2SLS model | 0.098 | 0.093 | 0.102 | 0.035 | 0.158 |
| Spatial Durbin model | 0.093 | 0.092 | 0.100 | 0.035 | 0.152 |

*, ** and *** denote statistical significance at the 10%, 5% and 1% level, respectively

## 7.2. Testing the instrument's strength

We apply the test of weak instruments. In order to test for this, we obtain the F-statistic for the instrumental variable in the first-stage regression of the 2SLS and compare it with the Stock-Yogo critical values. The F-statistic values in our 2SLS models as shown in Table 4 are shown in Table 7.[6] Apart from model (2) 2SLS (London's MSOAs), the F-statistics of all other models well exceed the rule of thumb suggested by Stock and Yogo (2005) that this value should be at least 10. A comparatively lower strength of the historical ADSL coverage as an instrument for download speeds in London can be explained by a low variance in ADSL coverage in that area. By 2001, ADSL deployment in London was nearly complete, as seen on the map in Figure 3. This results in a relatively low variation in the instrumental variable in London's subsample, as the vast majority of the localities reached universal ADSL coverage by 2001. However, for areas other than London, the strength of the instrument supports the validity of our 2SLS estimates. This suggests that the results from Table 4 should be selected instead of those in Table 3. Complete results of first-stage regression from the 2SLS models are presented in Table A3 in the Appendix.

---

[6] The F-statistic refers to the joint F-statistic for significance of the set of instruments. In our case, we have only one IV, hence the value of the F-statistic in our case is equivalent to the square of T-statistic of the instrument in the first-stage equation, as shown in Table A3. The F-statistics for the entire models are shown at the bottom of Table A3.

Table 7. Test for weak instrument, null hypothesis: weak IV

| Variables/Model | | (1) 2SLS | (2) 2SLS | (3) 2SLS | (4) 2SLS | (5) 2SLS | (6) 2SLS |
|---|---|---|---|---|---|---|---|
| F-statistic | | 48.1 | 4.77 | 46.6 | 23.2 | 27.5 | 28.9 |
| Stock-Yogo critical values | 30% | 1.07 | | | | | |
| | 10% | 2.71 | | | | | |
| | 5% | 3.84 | | | | | |
| | 1% | 6.63 | | | | | |

We estimate further test scores to examine whether our explanatory variable (download speeds) has an endogenous relationship with business establishment growth rates. The validity of results of such tests is dependent on the validity of our instrument, in particular the validity of the exclusion restriction which remains untestable in our just-identified case. Nevertheless, they provide a formal way of testing whether the 2SLS estimates from Table 4 should be preferred over the single-stage results from Table 3. Testing for endogeneity of download speeds in our just-identified case is equivalent to performing a Hausman test comparing the 2SLS with the single-stage estimates (Hayashi, 2000).

The result of the Hausman test is shown in Table 8. With the exception of model (2) for London-based subsamples, the difference in the coefficient estimates of download speed between the single-stage and 2SLS models is significant to at least the 10% level. In the context of weak instrument test, the likely bias in the 2SLS estimate for London invalidates the result of the Hausman test. For other areas, the Hausman test shows that the single-stage and 2SLS regressions in Table 3 and Table 4 respectively are statistically different. Under such conditions, the 2SLS estimate of the effect of broadband speeds on business growth should be preferred over single-stage model estimates. Nevertheless, the difference in the estimated size of the effect is relatively small between the models, and the overall conclusions on the positive impact of broadband on business creation remain similar regardless of the choice of the model.

Table 8. Hausman test of the models in Table 3 and Table 4 (single-stage versus 2SLS models), examining the difference in coefficients of independent variable (download speed)

|  | (1) Entire Sample | (2) London | (3) Outside London | (4) Low Human Capital | (5) Medium Human Capital | (6) High Human Capital |
|---|---|---|---|---|---|---|
| Coefficient difference | 0.0248*** (0.00523) | -0.103 (0.106) | 0.0283*** (0.00546) | 0.0351** (0.0162) | 0.0151*** (0.00602) | 0.00852* (0.00545) |

*, ** and *** denote statistical significance at the 10%, 5% and 1% level, respectively

*7.3. Sectoral classification of establishments*

Previous studies find that the benefits of broadband are particularly large for knowledge-intensive sectors (Mack, 2014) but less clear or even negative for other sectors such as manufacturing (Stephens et al., 2022). In some cases, the effect of broadband is only significant whenever the data is disaggregated into sectors (Mack and Grubesic, 2009). Using high (spatial) resolution data, our results demonstrate statistical significance without considering sectoral disaggregation. Hence, to increase the comprehensiveness of our results, we examine how the download speed-business creation relationship is affected when we disaggregate the establishment data into sectors. In order to check the sensitivity of our results to different industrial categorisations of business establishments, we divide establishments into several sectors and rerun model (1) using 2SLS.

We apply five industrial categories to the business establishment data. First, we divide them into knowledge intensive business services (KIBS) which consist of firms in the following industries: information and communication; finance and insurance; professional, scientific and technical; and education. Second, we separate out business establishments in manufacturing. Third, we have business establishments in the arts, entertainment and recreation services. Fourth, we have establishments in wholesale, transport and storage, which can be considered logistics firm. Last, we have establishments in retail services.

We regress the growth rate of business establishments in the above sectors with the same set of independent, control and instrumental variables, applying the 2SLS approach outlined by equation (3). The results are shown below in Table 9. The relationship between download speed and business is positive in all cases. The coefficients for arts, entertainment and recreation services, as well as for KIBS are quite large at 0.142 and 0.106, respectively. Manufacturing displays the weakest relationship at 0.0235 and is not statistically significant. For logistics and retail services firms, the coefficients (0.035 and 0.0424, respectively) are significant but less than that for establishments in all industries, as shown in model (1) of Table 4 (0.0574).

These results are in line with that of the existing literature, where the largest benefits accrue to firms in industries that can directly make use of broadband functionalities. KIBS firms are able to leverage high Internet speeds to enhance knowledge sharing (Doloreux and Laperrière, 2014) while HD streaming enabled by faster broadband is conducive to the growth of entertainment firms. Logistics and retail services firms do not seem to require the functionalities of faster broadband for their operations with slower speeds likely being sufficient to meet their demands. For manufacturing firms, the benefits of faster broadband are even less evident.

Table 9. Regression results with business establishments divided by sector.

| Variable/Model | Knowledge Intensive Business Services (KIBS) | Manufacturing | Arts, entertainment and recreation services | Wholesale, transport and storage (logistics) | Retail services |
|---|---|---|---|---|---|
| Download speed | 0.106*** | 0.0235 | 0.142*** | 0.0350** | 0.0424*** |
|  | (0.00890) | (0.0182) | (0.0127) | (0.0163) | (0.00928) |
| Population | -0.619*** | -0.605*** | -0.681*** | -0.573*** | -0.671*** |
|  | (0.0754) | (0.143) | (0.107) | (0.110) | (0.0829) |
| Road density | -0.00364 | -0.00551 | -0.0249 | -0.0496* | -0.0214 |
|  | (0.0195) | (0.0425) | (0.0207) | (0.0284) | (0.0184) |
| Inverse time to town centre | -0.0118 | 0.0332 | 0.0100 | -0.00961 | -0.0122 |
|  | (0.0113) | (0.0251) | (0.0162) | (0.0220) | (0.0118) |
| Inverse time to empl hub | 0.0298*** | 0.0163 | 0.0355*** | 0.0465*** | 0.0134*** |
|  | (0.00432) | (0.0117) | (0.00672) | (0.00943) | (0.00500) |
| Inverse time to supermarket | 0.00920 | -0.0180 | 0.000402 | -0.0198 | 0.0105 |
|  | (0.0106) | (0.0242) | (0.0149) | (0.0204) | (0.0106) |
| Crime rate | -0.0271*** | 0.0541** | -0.0166 | 0.0182 | -0.0155 |
|  | (0.0103) | (0.0248) | (0.0154) | (0.0203) | (0.0114) |
| Labour cost | -1.417*** | -0.186 | -1.701*** | -0.0975 | -0.437*** |
|  | (0.117) | (0.267) | (0.174) | (0.201) | (0.128) |
| HH Index | -0.110 | -0.0118 | -0.313*** | -0.139 | -0.132* |
|  | (0.0675) | (0.132) | (0.0896) | (0.108) | (0.0705) |
| Population density | 0.0408 | -0.0278 | 0.0131 | 0.0774 | -0.270*** |
|  | (0.0951) | (0.180) | (0.115) | (0.139) | (0.0958) |
| Unemployment | -1.158*** | 0.621 | -0.749 | -2.772*** | 0.224 |
|  | (0.443) | (0.739) | (0.516) | (0.655) | (0.414) |
| % change in estab, lagged 1 | -0.352*** | -0.417*** | -0.367*** | -0.392*** | -0.400*** |
|  | (0.0194) | (0.0110) | (0.0123) | (0.0129) | (0.0124) |
| Constant | 14.12*** | 6.504*** | 16.29*** | 5.215*** | 9.280*** |
|  | (0.951) | (2.035) | (1.395) | (1.565) | (1.055) |
| N | 27,164 | 27,164 | 27,164 | 27,164 | 27,164 |

*, ** and *** denote statistical significance at the 10%, 5% and 1% level, respectively

*7.4. Effect of broadband speeds in low, medium and high human capital regions*

We apply a dummy variable test to see whether models using different subsets of the data for low, medium and high human capital areas have significantly different coefficients in terms of download speed and business creation. This approach follows Gujarati (1970) which is a variant of the Chow test (Chow, 1960). The Chow test is commonly used to determine structural breaks in time series data between different time periods but can also be applied to examine breaks in the stratification of our data between different human capital levels. This is achieved by modifying equation (3) as below:

(4) $\%change_{estab_{it}} = \alpha_0 + \alpha_i + \hat{\beta}\ln(\widehat{download}_{speed_{it}}) +$

$\gamma X_{it} + \theta\%change_{estab_{it-1}} + H + \hat{\mu}H\ln(\widehat{download}_{speed_{it}}) + \varepsilon_{it}$

Where $H$ refers to the dummy variable, for whether an area is low, medium or high in human capital. If the coefficient $\hat{\mu}$ is significant, then it suggests that the relationship between download speed and business creation are statistically different across the dummy categories. When comparing two categories, we remove the third. For example, when comparing $\hat{\mu}$ between low and high human capital areas, we remove observations for medium human capital areas. $H$ then takes on a value of 1 in equation (4) if an area has high human capital. This compares how high human capital areas differ from low human capital areas, removing the effect from medium human capital areas.

We apply 2SLS regression to equation (4) above. As download speed is endogenous, its multiplicative term by the dummy variable is also potentially endogenous. In order to address two endogenous variables, we generate a second corresponding instrument by multiplying our IV by the dummy variable. With two IVs and two endogenous variables, the 2SLS is again just-identified. The result of the analysis is shown in Table 10. The coefficient for low human capital area is significantly higher than that for medium and high human capital areas, but the coefficient for high human capital areas is not significantly higher than for medium human capital areas.

TABLE 10. Dummy Chow test applied to models using different levels of human capital, comparing the coefficient of the interaction term between human capital group dummy and download speeds, $\hat{\mu}$ in equation (4)

|  | Low | Medium | High |
|---|---|---|---|
| Low | - | 0.0214** (0.00918) | 0.0296*** (0.00985) |
| Medium | -0.0214** (0.00918) | - | 0.00470 (0.00388) |
| High | -0.0296*** (0.00985) | -0.00470 (0.00388) | - |

*, ** and *** denote statistical significance at the 10%, 5% and 1% level, respectively

## 8. DISCUSSION AND CONCLUSION

In this article the empirical relationship between NGA broadband and business creation is examined at the local level over the 2011-2015 period in England, United Kingdom. We find that faster broadband speeds brought by NGA technologies had a positive effect on the rate of business growth. The primary hypothesised mechanism behind the estimated relationship is the enabling effect of faster broadband speeds. Faster broadband speeds brought by NGA deployment enable new business models based on digital technologies. Entrepreneurs either sought appropriate locations that offer high quality broadband infrastructure (contributing to new business establishment growth), or potentially enjoyed a competitive advantage (resulting in a higher survival rate).

The headline estimates in this study are obtained using two-stage-least-squares (2SLS) panel data fixed-effects modes. Relying on the instrumental variable (IV) approach allows us to account for potential endogeneity running from business creation to broadband speed. We find that in England on average a one percentage increase in download speeds is associated with a 0.0574 percentage point higher annual growth rate in the number of businesses. To put this into perspective in terms of averaged values, an average MSOA in England had 326 business establishments in 2011, which grew to 375 in 2015. During this period, the average broadband speed went from 6.3 Mbps in 2011 to 21.1 Mbps in 2015. Hypothetically, had broadband speed remained at the 2011 level, then we can apply a reduction to the annual growth rate in establishment equal to the percentage increase in download speed multiplied by the coefficient of 0.0574. This represents a counterfactual scenario in which broadband infrastructure assets were not improved and the resulting lower growth in establishment can be interpreted as the growth attributed to the increase in download speed. In Figure 4 below we plot the average number of establishments against this hypothetical scenario where average download speed remained unchanged. The increase in average establishment each year attributed to the increase in download speed is also shown.

Figure 4. Average number of establishments in MSOA 2011-2015, with and without increase in download speed

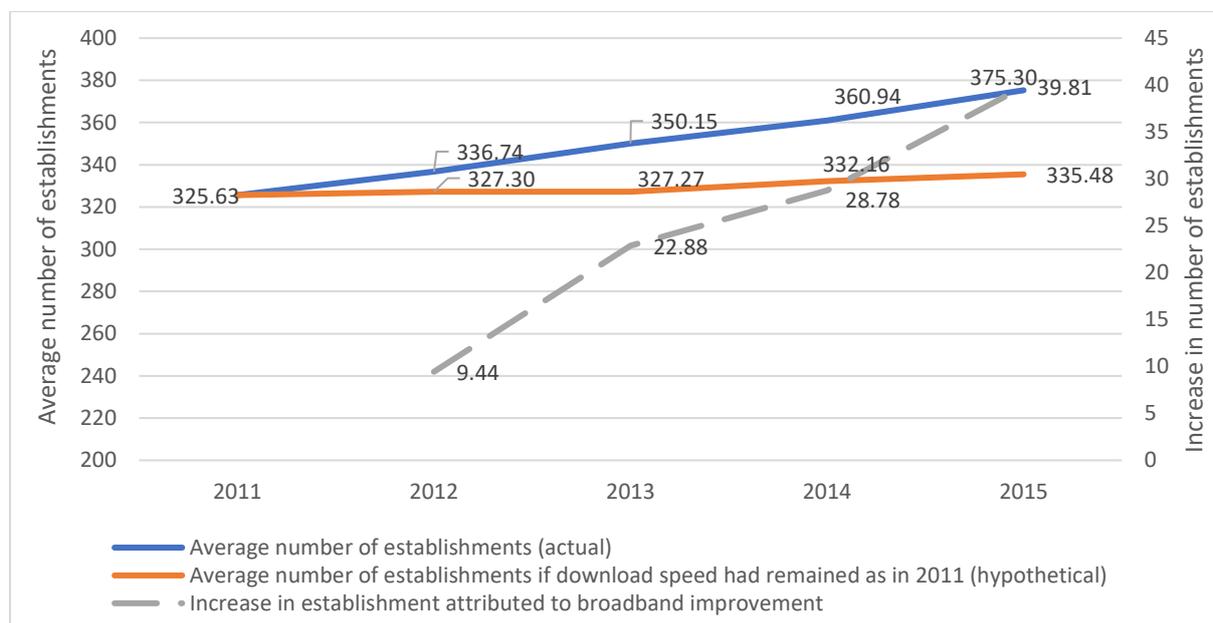

Without upgrading broadband infrastructure assets, by 2015 the average MSOA would have almost 40 fewer establishments. At an estimated 335 establishments on average in 2015, it would only be slightly above the 2011 value of 326 had download speed not been increased, which implies that the business creation effect of broadband is substantial.

Evidence from sub-sample analysis suggests that this effect could have been even higher in non-metropolitan areas since the estimated coefficients are larger when London is excluded from our analysis. As our instrument is a relatively weak predictor of download speeds in London, we are not able to identify a specific estimate of the size of the effect in that model.

Previous studies such as McCoy et al. (2018) exclude highly metropolitan areas from the analysis, since these regions are likely to respond differently to broadband provision, often relying on more dedicated lines and infrastructure. We find that the effects of download speed on business creation are largest in low human capital areas, followed by high human capital areas. This suggests that NGA broadband may assist low human capital regions in generating new types of businesses, while high human capital areas can also exploit many of the applications brought by the new type of broadband technology. On the other hand, areas with medium levels of human capital seem to benefit less, though the difference in coefficient with high human capital areas is not statistically significant. Therefore, our results show some indication that the effect of broadband on local business creation is moderated by human capital.

Our results are robust to a number of alternative model specifications, including restricting the estimation sample to localities with a sufficient number of crowdsourced datapoints; the analysis in level forms; the inclusion of lagged independent variable; and the application of the spatial Durbin model. The instrument we select, lagged ADSL coverage, is a strong predictor of download speeds in all sub-samples except when examining London-only areas. Testing for endogeneity using the Hausman test confirms that it is preferable to rely on 2SLS models due to the possibility that the relationship between download speed and business creation is endogenous. The effects of broadband on business creation appear to be different across establishments in different industries. We find firms in knowledge-intensive business services (KIBS) and arts, entertainment and recreation sectors displaying a particularly strong download speed-business creation relationship.

Our study supports the general conclusion of positive economic impact from faster broadband speeds brought by NGA. Nonetheless, there are several limitations to this study that ought to be pointed out. Firstly, our data covers a relatively short period of only five years, mostly during or just after the main phase of NGA broadband deployment. It is possible that for many areas which received the deployment before the studied period, a part of the positive economic effects has already materialised. Alternatively, it can also be hypothesised that faster broadband speeds continue to affect the business establishment decisions well into the future, beyond the studied period. This is because growth may only emerge later when both the technology and their business model applications become sufficiently mature. This is a well-known problem in economics related to ICT exhibiting a 'productivity paradox' where productivity growth materialises only when 'killer applications' for these technologies emerge, perhaps many years later. As more recent data become available, researchers could investigate the profile of impacts from new broadband technologies, for example by estimating impulse response functions. Secondly, our data which is based on the average broadband speed observed within an MSOA obtained mostly from residential locations, may not fully represent the actual download speed experienced by any specific business. Some businesses may choose to purchase premium broadband services that are considerably faster than typical residential speeds, hence there may be substantial variance of realised download speed at the firm level. This problem is likely to be particularly acute in the case of London, but for other areas of England it can also be the case, albeit to a lesser degree. The third limitation is the potential sensitivity of the results to the choice of the instrument. It is an extremely challenging task to compile high-resolution data over time for suitable instruments, which limits out options. While the choice of historical ADSL coverage is supported by logical reasoning similar to Duranton and Turner (2012), and augmented by testing for the strength of our instrument, the exclusion restriction cannot be tested with just a single variable.

There are several policy implications from these findings. Over the time period assessed in this analysis many governments around the world have pursued policies that aim to enhance the coverage and capacity of broadband infrastructure particularly in rural and remote regions. The findings of this study suggest that aspiring to reach near-universal broadband connectivity and dramatically increase download speed that would enable qualitatively different types of business applications may indeed be economically beneficial, especially as the costs of delivering such service decline. The availability of broadband can enable new business models serving customers domestically and abroad, causing an increase in business activities. The returns to broadband investments are likely to be larger in low human capital area which may benefit from a catch-up effect, as well as in high human capital areas with the complementary skills to exploit these applications.

This study which employed particularly high-resolution data, shows that these dynamics can have an impact at the very local level and that broadband infrastructure matters for business location decisions. It is beyond the scope of this study to explore other possible economic benefits of faster broadband speeds brought by NGA, such as increased employment or output. Other studies which focus on these effects point to there being positive impacts (Koutroumpis, 2009; Czernich et al., 2011), and these can be fruitful avenues for future research. However, future studies should seek to analyse the benefits of broadband at the very local level, as well as focusing on more advanced forms of broadband infrastructure.

# APPENDIX

TABLE A1: Summary of data

| Variable | Type | Description | Source |
| --- | --- | --- | --- |

| Growth rate of business establishment | Dependent variable | Number of business establishments, broken down by size and sector | ONS Nomis (2018): UK Business Counts |
|---|---|---|---|
| Broadband percentage coverage | Independent variable | Observed download speeds averaged across MSOAs | Broadband Speed Checker (2021) for 2011-2015 |
| Road density | Control variable | Length of major road per unit area. Major roads includes all principle and trunk roads as defined by the DfT. | DfT (2011, 2018): compiled from traffic count points |
| Inverse travel time to (nearest) town centre | Control variable | The inverse travel time to the nearest town centre by walk or public transport. Denotes centrality. | DfT (2011, 2018)[7] accessibility statistic, 2014 and 2015 adjusted for comparability with 2011-13 |
| Inverse travel time to (nearest) centre of employment | Control variable | The inverse travel time to the nearest employment centre. An employment centre is a settlement with at least 100 jobs. | DfT (2011, 2018): accessibility statistic |
| Inverse travel time to (nearest) supermarket | Control variable | The inverse travel time to the nearest supermarket by walk or public transport. Proxy for local amenity and services. | DfT (2011, 2018): accessibility statistic |
| Proportion of workforce with tertiary degree | Control variable through stratification | The percentage of workforce with higher education degree – an indication of the human capital or quality of local workforce. | Census for 2011 (Census, 2011) |
| Population density | Control variable | Population per unit area | Local population estimates from ONS |
| Unemployment rate | Control variable | Unemployment rate by jobseekers' allowance. Reflects the level of local factor demand. | ONS Nomis (2018): UK Business Register and Employment Survey |
| Crime and antisocial behaviour per 1000 | Control variable | Shows the crime rate in an area. Denotes the level of deprivation and poor environment, which should discourage business creation ceteris paribus. | Data.Police (2020): UKCrimeStats, a data platform from the Economic Policy Centre |
| Labour cost | Control variable | Index of labour cost per hour. Another indication of local demand. Seasonally adjusted. | ONS Nomis (2018): ONS estimate of labour cost per hour seasonally adjusted, divided by industrial sector. Employment proportion in each sector used to compute local labour cost index. |
| HH index | Control variable | Spatial Herfindahl-Hirschman index measuring the degree of industrial specialisation. | Computed using local share of employment in each industry. |
| Total employment | Control variable | Total number of people employed | ONS Nomis (2018): UK Business Register and Employment Survey |
| ADSL coverage | Instrumental variable | The percentage of premises in MSOA covered by ADSL lagged | SamKnows(2020) |

TABLE A2: Variable descriptive statistics

| Variable | Year | Unit | Mean | Std. Dev. | Min | Max |
|---|---|---|---|---|---|---|
| Growth rate in business establishment | 2011[8] | % change | 0.611 | 6.55 | -51.95 | 272 |
|  | 2015 | % change | 3.97 | 9.44 | -90.43 | 182 |
|  | 2011-2015 | 2011-2015 change | 3.36 | 11.49 | -267 | 187 |
| Download speed | 2011 | Mbps (natural log used in regression) | 6.30 | 3.70 | 0.715 | 38.37 |

---

[7] The DfT slightly altered the method of calculating accessibility in 2013. In order to make it as comparable as possible with the data in pre-2013, we adjusted the data in 2014 and 2015 using a scale factor based on the change in 2013-2014, applied to all MSOAs in 2014 and 2015.

[8] Establishment data for 2010 is used to compute the growth rate of establishment in 2011.

| | | | | | | |
|---|---|---|---|---|---|---|
| | 2015 | Mbps (natural log used in regression) | 21.07 | 6.43 | 4.53 | 73.10 |
| | 2011-2015 | 2011-2015 change | 14.78 | 6.31 | -19.57 | 67.70 |
| Population | 2011 | Number of people (natural log used in regression) | 7,820 | 1,612 | 2,224 | 16,439 |
| | 2015 | Number of people (natural log used in regression) | 8,067 | 1,778 | 2,324 | 18,436 |
| | 2011-2015 | 2011-2015 change | 247 | 415 | -1,222 | 8,879 |
| Road density | 2011 | Km/ hectare (natural log used in regression) | 0.430 | 0.621 | 0 | 6.71 |
| | 2015 | Km/ hectare (natural log used in regression) | 0.451 | 0.625 | 0 | 6.71 |
| | 2011-2015 | 2011-2015 change | 0.0207 | 0.106 | 0 | 1.91 |
| Inverse time to town centre | 2011 | Minutes inversed (natural log used in regression) | 8.23 | 3.49 | 0.833 | 20.00 |
| | 2015 | Minutes inversed (natural log used in regression) | 8.16 | 3.59 | 0.833 | 34.08 |
| | 2011-2015 | 2011-2015 change | -0.0649 | 1.53 | -10.17 | 14.08 |
| Inverse time to (nearest) employment hub | 2011 | Minutes inversed (natural log used in regression) | 13.13 | 4.03 | 2.21 | 20.00 |
| | 2015 | Minutes inversed (natural log used in regression) | 15.45 | 7.08 | 0.984 | 56.33 |
| | 2011-2015 | 2011-2015 change | 2.32 | 6.83 | -14.13 | 37.35 |
| Inverse time to (nearest) supermarket | 2011 | Minutes inversed (natural log used in regression) | 15.48 | 3.61 | 0.833 | 20.00 |
| | 2015 | Minutes inversed (natural log used in regression) | 16.62 | 6.55 | 0.833 | 50.10 |
| | 2011-2015 | 2011-2015 change | 11.41 | 4.29 | -9.27 | 30.10 |
| Crime rate | 2011 | Number per 1,000 inhabitants (natural log used in regression) | 14.47 | 12.99 | 1.78 | 354.50 |
| | 2015 | Number per 1,000 inhabitants (natural log used in regression) | 13.81 | 11.80 | 1.96 | 305.11 |
| | 2011-2015 | 2011-2015 change | -6.64 | 3.87 | -113.49 | 48.77 |
| Labour cost | 2011 | Index (natural log used in regression) | 465.09 | 16.85 | 350.38 | 488.21 |
| | 2015 | Index (natural log used in regression) | 503.67 | 16.67 | 383.07 | 531.22 |
| | 2011-2015 | 2011-2015 change | 38.59 | 6.91 | -39.91 | 106.31 |
| Herfindahl-Hirschman index | 2011 | Index (range 0-1) | 0.184 | 0.0931 | 0.0782 | 0.837 |
| | 2015 | Index (range 0-1) | 0.179 | 0.0905 | 0.0748 | 0.864 |
| | 2011-2015 | 2011-2015 change | -0.0524 | 0.0482 | -0.384 | 0.474 |
| Unemployment rate | 2011 | Percent (range 0-100%) | 2.44 | 1.71 | 0.246 | 18.42 |
| | 2015 | Percent (range 0-100%) | 0.892 | 0.758 | 0.0413 | 5.98 |
| | 2011-2015 | 2011-2015 change (range 0-100%) | -1.55 | 1.09 | -16.59 | 0.845 |
| Proportion of workforce with tertiary degree | 2011 | Percent (range 0-100%) | 27.28 | 8.97 | 1.62 | 71.60 |
| | 2015 | Percent (range 0-100%) | 29.48 | 10.45 | 0 | 86.93 |
| | 2011-2015 | 2011-2015 change (range 0-100%) | 2.20 | 2.34 | -9.58 | 15.34 |
| Coverage of ADSL broadband | 2001 | Percent (range 0-100%) | 66.54 | 44.49 | 0 | 100 |
| | 2005 | Percent (range 0-100%) | 99.91 | 1.66 | 10 | 100 |
| | 2001-2005 | 2001-2005 change (range 0-100%) | 33.37 | 44.39 | 0 | 100 |

TABLE A3: First-stage regression results of 2SLS models in Table 4

| Variable/Model | (1) 2SLS Entire Sample | (2) 2SLS London | (3) 2SLS Outside London | (4) 2SLS Low Human Capital | (5) 2SLS Medium Human Capital | (6) 2SLS High Human Capital |
|---|---|---|---|---|---|---|
| IV (ADSL coverage 2001-2005) | 0.539*** | 0.563*** | 0.532*** | 0.383*** | 0.509*** | 0.716*** |
|  | (0.0112) | (0.118) | (0.0114) | (0.0165) | (0.0185) | (0.0248) |
| Population | 0.658*** | 1.271*** | 0.513** | 0.417 | -0.183 | 0.479* |
|  | (0.224) | (0.334) | (0.225) | (0.282) | (0.373) | (0.281) |
| Road density | 0.204*** | 0.0617 | 0.217*** | 0.107 | 0.188** | 0.260*** |
|  | (0.0484) | (0.104) | (0.0516) | (0.0708) | (0.0734) | (0.0937) |
| Inverse time to town centre | 0.114*** | -0.275*** | 0.156*** | 0.106*** | 0.161*** | 0.0349 |
|  | (0.0244) | (0.0703) | (0.0255) | (0.0382) | (0.0374) | (0.0480) |
| Inverse time to employment hub | 0.0504*** | 0.0638*** | 0.0436*** | 0.0528*** | 0.0113 | 0.0606*** |
|  | (0.0114) | (0.0238) | (0.0122) | (0.0152) | (0.0176) | (0.0208) |
| Inverse time to supermarket | -0.256*** | 0.0245 | -0.320*** | -0.112*** | -0.359*** | -0.220*** |
|  | (0.0210) | (0.0423) | (0.0236) | (0.0360) | (0.0350) | (0.0390) |
| Crime rate | -0.0212 | -0.442*** | 0.0244 | -0.0310 | 0.00539 | -0.0544 |
|  | (0.0232) | (0.0714) | (0.0248) | (0.0397) | (0.0386) | (0.0408) |
| Labour cost | 2.888*** | 1.152** | 2.992*** | 1.011*** | 2.591*** | 4.940*** |
|  | (0.251) | (0.534) | (0.269) | (0.288) | (0.398) | (0.637) |
| HH Index | -0.412*** | -0.0758 | -0.463*** | -0.308** | -0.623*** | -0.365** |
|  | (0.0957) | (0.185) | (0.107) | (0.131) | (0.202) | (0.183) |
| Population density | 1.965*** | 0.473 | 2.241*** | 1.454*** | 1.352*** | 1.741*** |
|  | (0.216) | (0.371) | (0.224) | (0.281) | (0.380) | (0.429) |
| Unemployment | -19.75*** | -21.63*** | -19.47*** | -19.52*** | -29.84*** | -18.82*** |
|  | (0.775) | (1.568) | (0.815) | (0.657) | (2.339) | (2.996) |
| % change in estab, lagged 1 | 0.162 | 0.115* | 0.163 | 0.00535 | 0.484*** | 0.903*** |
|  | (0.101) | (0.0638) | (0.111) | (0.0359) | (0.0907) | (0.302) |
| Constant | -25.50*** | -16.89*** | -25.11*** | -10.70*** | -14.21*** | -36.11*** |
|  | (2.193) | (3.755) | (2.308) | (2.874) | (3.847) | (4.691) |
| N | 27,164 | 3,932 | 23,232 | 10,724 | 8,560 | 7,880 |
| Model F-statistic (IV included in regression) | 892 | 111 | 849 | 337 | 429 | 336 |
| Model F-statistic (IV excluded in regression) | 417 | 117 | 386 | 242 | 213 | 98 |

*, ** and *** denote statistical significance at the 10%, 5% and 1% level, respectively